\documentclass[aps,prc,superscriptaddress,twoside,twocolumn,nofootinbib,%
showpacs,floatfix,dvips]{revtex4}

\usepackage{amsmath,amssymb,graphicx,bm}

\begin{document}

%%%%%%%%%%%%%%%%%%%%% Title %%%%%%%%%%%%%%%%%%%%%%

\title{\boldmath
Heavy baryon/meson ratios in relativistic heavy ion collisions}

%%%%%%%%%%%%%%%%%%%% Authors %%%%%%%%%%%%%%%%%%%%%
%%%%%%%%%%%%%%%%%%%% Addresses %%%%%%%%%%%%%%%%%%%%%

\author{Yongseok Oh}%
\email{yoh@comp.tamu.edu}

\affiliation{Cyclotron Institute and Physics Department, Texas A\&M
University, College Station, Texas 77843, U.S.A.}

\author{Che Ming Ko}%
\email{ko@comp.tamu.edu}

\affiliation{Cyclotron Institute and Physics Department,
Texas A\&M University, College Station, Texas 77843, U.S.A.}

\author{Su Houng Lee}%
\email{suhoung@phya.yonsei.ac.kr}

\affiliation{Institute of Physics and Applied Physics, Yonsei
University, Seoul 120-749, Korea}

\author{Shigehiro Yasui}%
\email{yasuis@post.kek.jp}

\affiliation{
High Energy Accelerator Research Organization (KEK), Tsukuba,
Ibaraki, 302-0801, Japan}

\date{\today}

%%%%%%%%%%%%%%%%%%%% Abstract %%%%%%%%%%%%%%%%%%%%%

\begin{abstract}

Heavy baryon/meson ratios $\Lambda_c/D^0$ and $\Lambda_b/\bar{B}^0$
in relativistic heavy ion collisions are studied in the quark
coalescence model. For heavy baryons, we include production from
coalescence of heavy quarks with free light quarks as well as with
bounded light diquarks that might exist in the strongly coupled
quark-gluon plasma produced in these collisions. Including the
contribution from decays of heavy hadron resonances and also that
due to fragmentation of heavy quarks that are left in the system
after coalescence, the resulting $\Lambda_c/D^0$ and
$\Lambda_b/\bar{B}^0$ ratios in midrapidity ($|y|\le 0.5$)
from central Au+Au collisions at $\sqrt{s_{NN}^{}}=200$~GeV are
about a factor of five and ten, respectively, larger than those
given by the thermal model, and about a factor of ten and twelve,
respectively, larger than corresponding ratios in the PYTHIA model
for $pp$ collisions. These ratios are reduced by a factor of about
1.6 if there are no diquarks in the quark-gluon plasma. The
transverse momentum dependence of the heavy baryon/meson ratios is
found to be sensitive to the heavy quark mass, with the
$\Lambda_b/\bar{B}^0$ ratio being much flatter than the
$\Lambda_c/D^0$ ratio. The latter peaks at the transverse momentum
$p_T^{} \simeq 0.8$~GeV but the peak shifts to $p_T^{} \simeq 2$~GeV
in the absence of diquarks.

\end{abstract}

\pacs{
      25.75.-q, % relativistic heavy ion collisions
      25.75.Cj, % photon, lepton, and heavy quark production in rhic
      25.75.Dw, % particle and resonance production
     }

\maketitle

\section{Introduction}

Recent measurements on nonphotonic electrons from decays of
midrapidity heavy-flavored mesons in central heavy ion
collisions at the Relativistic Heavy Ion Collider (RHIC) have
shown that the nuclear modification factor $R_{AA}^e$ relative to
$pp$ collisions is significantly smaller than theoretical
predictions based on radiative energy loss of heavy quarks in
produced quark-gluon plasma~\cite{PHENIX06b,STAR06c,DK01b,DG04}. To
explain the observed small value of $R_{AA}^e$ or large suppression
of heavy meson production in relativistic heavy ion collisions,
other mechanisms for energy loss of heavy quarks in quark-gluon
plasma have been proposed~\cite{VR04,DGVW05,VGR06,KL07,ZCK05,GAP08}.
Also, it was suggested that the enhancement of charmed baryon
production over charmed meson production, i.e., the $\Lambda_c/D^0$
ratio, would suppress the $R_{AA}^{e}$ of the electrons from heavy
mesons~\cite{SD05-Soren07,MGC07}. This is due to the fact that
enhanced charmed baryon production reduces the production of charmed
mesons and thus their contribution to decay electrons.

In fact, experiments at RHIC have shown that in central heavy
ion collisions there is an enhanced production of midrapidity baryons in
the intermediate transverse momentum region not only in light
hadrons but also in strange hadrons. This enhancement could be
described by multi-quark dynamics through quark coalescence or
recombination~\cite{GKL03,GKL03b,FMNB03}, baryon junction
loops~\cite{Khar96,VGW98,VG99}, or long-range coherent
fields~\cite{TGBG05}. Therefore, one might expect enhanced heavy
baryon production in the intermediate transverse momentum region as
well. Since the $\Lambda_c/D^0$ ratio has not been measured in heavy
ion collisions, it was assumed in Ref.~\cite{SD05-Soren07} to be the
same as the measured $\Lambda/K_S^0$ ratio or in Ref.~\cite{MGC07}
to have a Gaussian form with a maximum value peaked at the
transverse momentum $p_T^{} = 5$~GeV. It was then claimed that a
value of $\Lambda_c/D^0 \sim 1$, which is about a factor of ten
larger than that in $pp$ collisions at same energy, could result in
an additional $20$--$25$\% suppression of $R_{AA}^e$ apart from the
suppression due to the collisional energy loss of heavy quarks in
quark-gluon plasma. However, it is not yet known what kind of
underlying physics can cause such a large enhancement of the
$\Lambda_c/D^0$ ratio.

Recently, enhancement of the $\Lambda_c/D^0$ ratio due to the
existence of $[ud]$ diquarks in strongly coupled quark-gluon plasma
(QGP) was suggested by Lee {\it et al.}~\cite{LOYYK07}. This was
based on the idea of possible existence of quasi bound states of
quarks and gluons including diquarks in QGP when the temperature is
between $T_C^{}$ and $4T_C^{}$~\cite{SZ03b-SZ04}, where $T_C^{}$ is
the critical temperature for the quark-gluon plasma to hadronic
matter transition. Based on the quark coalescence model, it was
shown in Ref.~\cite{LOYYK07} that the $\Lambda_c/D^0$ ratio could be
enhanced by a factor of $4$--$8$ relative to the case without
diquarks in QGP, depending on the binding energy of the light
diquark at $T_C^{} = 175$~MeV. The question whether diquarks can
exist as quasi bound states in cold quark matter was raised earlier
in Ref.~\cite{DS88}, where the quark-diquark matter was claimed to
be energetically more favorable than the free quark matter. The idea
of enhancement of $\Lambda_c$ baryon production in heavy-ion
collisions was then suggested in Ref.~\cite{Sateesh92}. Based on the
assumption that the number of diquarks present in a collision is
mainly determined by that in the initial nuclei, the author of
Ref.~\cite{Sateesh92} considered the $\Lambda_c/\Sigma_c$ ratio and
estimated that it could be enhanced by a factor of as large as $80$.
Since the $\Sigma_c$ can not be directly measured in near-future
heavy-ion experiments, the more phenomenologically accessible
$\Lambda_c/D^0$ ratio is thus addressed in Ref.~\cite{LOYYK07}.

The idea of diquarks~\cite{IK66,LT67-LTK68,SV05} has been widely
used in hadron physics for describing the mass spectrum,
electromagnetic properties, and many other properties of hadrons
(for a review, see, for example, Ref.~\cite{APEFL93}), although the
properties and even the existence of diquarks are still under
debate~\cite{BDTR02}. Among possible signatures of diquarks in
hadrons, the observed large ratio of $\Lambda_c/\Sigma_c \sim 7$ in
$e^+e^-$ collisions was pointed out in Ref.~\cite{SW06}, which
implies a significantly larger production of scalar diquarks than
vector diquarks~\cite{EFJL83}. (See also Ref.~\cite{Jaffe04}.) In
spite of the lack of concrete consensus on the diquark nature, it is
well-known that the (isoscalar) scalar light $[ud]$ diquark of color
anti-triplet is the most probable diquark state. Evidence for such
scalar diquarks has been confirmed by a recent lattice QCD
simulation~\cite{ADL06}. (See, however, Ref.~\cite{HKLW98} for other
lattice QCD calculation on diquarks.)

In diquark models, the $\Lambda_c$ is usually considered as a system
consisting of a charm quark and a scalar light $[ud]$ diquark. On
the other hand, the structure of $\Sigma_c$ and $\Sigma_c^*$ is
model-dependent: they may have an axial-vector light diquark but
they also may contain diquarks $[Qq]$ made of one heavy quark ($Q$)
and one light quark ($q$)~\cite{LNW83,EFKR96}. In Ref.~\cite{FSR88},
it is claimed that the ground state baryons favor the $[Qq]$
diquark. This is based on the observation that the $[Qq]q$ system
gives a $\Lambda_c$ mass closer to that from the three-quark
calculation, although it has a larger mass than that based on the
$Q[qq]$ configuration. Since the color-spin interaction between
quarks is inversely proportional to the quark masses, the binding
energies of $[Qq]$ diquarks are smaller than those of scalar light
$[qq]$ diquarks, and it is not clear whether such $[Qq]$ diquarks
can exist, in particular, in QGP. (See Ref.~\cite{HNV08} for a
recent discussion on the diquark picture for heavy baryons.)

In this paper, we calculate the $\Lambda_Q/H^0$ ratios ($Q=b,c$ and
$H^0$ stands for $D^0$ or $\bar{B}^0$) in relativistic heavy ion
collisions based on the quark coalescence model. We consider the two
cases of $\Lambda_Q$ being a pure three-quark state and of
$\Lambda_Q$ being made of a heavy quark and one scalar light
diquark. In addition to the direct formation of $\Lambda_Q$ and
$H^0$, we also take into account the contribution from decays of
heavy hadron resonances. Namely, we consider the decay of $D^*$
mesons for $D^0$ production and $\Sigma_Q$ and $\Sigma_Q^*$ baryon
decays for $\Lambda_Q$ production. As we have discussed before, we
assume that $\Sigma_Q$ and $\Sigma_Q^*$ baryons are made of three
quarks and that neither vector light diquarks nor scalar heavy-light
diquark exists in QGP.

This paper is organized as follows. In the next section, we briefly
discuss heavy hadron production in $pp$ collisions. A simple thermal
model is then used in Sec.~\ref{sec:thermal} to show the role of
heavy hadron resonances in the $\Lambda_Q/H^0$ ratios. The
difference between the ratios for charmed hadrons and for bottom
hadrons is also discussed. In Sec.~\ref{sec:coal}, the coalescence
model used in this work is given together with both light and heavy
quark distribution functions. Also described are the calculational
methods and the contributions from fragmentation of heavy quarks
that are not used in coalescence. We then show in the same section
the results for the relative production fractions of heavy hadrons
and their $p_T^{}$ spectra as well as the transverse momentum
dependence of the $\Lambda_Q/H^0$ ratios. Section~\ref{sec:summary}
contains the conclusions and discussions.

\section{Heavy hadron production in proton-proton collisions}
\label{sec:pp}

Before studying the production of heavy hadrons in heavy ion
collisions, we briefly discuss their production in $pp$ collisions
by using the PYTHIA model~\cite{SMS06}. For $pp$ collisions at
$\sqrt{s} = 200$~GeV, the PYTHIA model gives following relative
production fractions of charmed hadrons in midrapidity
($|y|\le 0.5$),%
\footnote{We have confirmed the results reported in
Ref.~\cite{MGC07}. These values are also close to the values quoted
by Particle Data Group (PDG)~\cite{PDG08} for $e^+e^-$ annihilation
at $\sqrt{s} = 91$~GeV: $f(D^+) \simeq 0.21$, $f(D^0) \simeq 0.54$,
$f(D_s^+) \simeq 0.16$, and $f(\Lambda_c) \simeq 0.093$.}

\begin{eqnarray}
&& f(D^+) \simeq 0.196, \qquad f(D^0) \simeq 0.607, \qquad
\nonumber \\ &&
f(D_s^+) \simeq 0.121, \qquad f(\Lambda_c) \simeq 0.076.
\label{eq:PYTHIA}
\end{eqnarray}
This leads to the particle number ratios
\begin{eqnarray}
\left(\frac{D^0}{D^+}\right)_{pp} \simeq 3.1, \qquad
\left(\frac{\Lambda_c}{D^0}\right)_{pp} \simeq 0.13.
\label{eq:PYTHIA2}
\end{eqnarray}

The above estimated $D^0/D^+$ ratio can be understood largely by
considering resonance decays, namely, decays of $D^*$ mesons into
$D$ mesons. First, the $D^{*0}$ meson decays into $D^0$ meson by
100\%~\cite{PDG08} since the decay of $D^{*0}$ into $D^+ \pi^-$ is
prohibited by energy conservation.%
\footnote{Note that $m(D^{*0}) = 2007$~MeV, $m(D^{*+}) = 2010$~MeV,
$m(D^0) = 1865$~MeV, $m(D^+) = 1869$~MeV, while $m(\pi^\pm) =
139.6$~MeV and $m(\pi^0) = 135$~MeV.} Second, the $D^{*+}$ meson can
decay either into $D^+$ or $D^0$. The branching ratios are
$BR(D^{*+} \to D^0 \pi^+) \approx 68$\% and $BR(D^{*+} \to D^+ X)
\approx 32$\%, where sum over any state $X$ is understood. If the
production rates are equal for $D$ and $D^*$ mesons aside from the
factor due to different spin degeneracies, we then would have
\begin{equation}
\left( \frac{D^0}{D^+} \right)_{pp} \approx
\frac{1 + 3 + 3 \times 0.68}{1 + 3 \times 0.32}
= \frac{6.04}{1.96} \approx 3.1,
\label{eq:D*decay}
\end{equation}
where the spin degeneracy of $D^*$ is included. We also note that
the estimated $\Lambda_c/D^0$ ratio is close to that ($\approx
0.16$) in the SELEX measurement~\cite{SELEX00}.

For bottom hadron production in $pp$ collisions at
$\sqrt{s_{NN}^{}}=200$~GeV, the relative production fractions in
midrapidity from the PYTHIA model are
\begin{eqnarray}
&& f(B^-) \simeq 0.101, \quad f(\bar{B}^0) \simeq 0.101,
\quad f(B_s^-) \simeq 0.030,
\nonumber \\
&&  f(B^{*-}) \simeq 0.302, \quad f(\bar{B}^{*0}) \simeq 0.301,
\quad f(B_s^{*-}) \simeq 0.089, \nonumber \\
&& f(\Lambda_b) \simeq 0.076,
\label{eq:PYTHIAb}
\end{eqnarray}
leading to%
\footnote{We note that the average fractions of bottom mesons and
bottom baryons in $p\bar{p}$ annihilation at $\sqrt{s} = 1.5$~TeV
has been estimated to be $f(b\bar{u}) = f(b\bar{d}) = 0.399$ and
$f({\rm baryon}) = 0.092$~\cite{PDG08}. If we assume that $b\bar{q}$
mesons include equal number of $B$ and $B^*$ mesons apart from the
spin degeneracy and the number of bottom baryons is mostly due to
$\Lambda_b$, then we would have $\Lambda_b/\bar{B}^0 \approx 1.1$ in
$p\bar p$ collisions.}
\begin{equation}
\Lambda_b/\bar{B}^0 \approx 0.7,
\label{eq:PYTHIA4}
\end{equation}
which is significantly larger than the $\Lambda_c/D^0$ ratio given
in Eq.~(\ref{eq:PYTHIA2}).

\section{The thermal model}
\label{sec:thermal}

In relativistic heavy-ion collisions, the $\Lambda_c/D^0$ and
$\Lambda_b/\bar{B}^0$ ratios are expected to be different from those
given in Eqs.~(\ref{eq:PYTHIA2}) and (\ref{eq:PYTHIA4}) for $pp$
collisions. This can be seen using a simple thermal model which
assumes that in relativistic heavy ion collisions charmed and
bottom hadrons are produced during hadronization of the
quark-gluon plasma and that they are in both thermal and chemical
equilibrium at the phase transition temperature $T_C$. The
total charm and bottom numbers are, however, determined by initial
hard scattering as thermal production of heavy quarks from created
quark-gluon plasma is negligible in heavy ion collisions at
RHIC~\cite{ZKL07}. The number of heavy hadrons of mass $m$ inside a
fireball of volume $V$ at $T_{C}$ is then
given by
\begin{equation}
N = \gamma_Q^{} \frac{gV}{2\pi^2} m^2 T K_2(m/T_{C}),
\end{equation}
where $\gamma_Q^{}$ is the heavy-quark fugacity, $g$ is the
degeneracy of the particle, and $K_2$ is the modified Bessel
function.

For charmed hadrons, we would then have $(D^0/D^+)_0 = 1$, where the
subscript $0$ means the ratio without resonance contributions, if
the $D^* \to D$ decay is not included. Since
\begin{equation}
\frac{D^{*0}}{D^0} = \frac{3 m_{D^*}^2 K_2(m_{D^*}/T_{C})}
{m_{D_0}^2 K_2(m_{D_0}/T_{C})} \simeq 1.47
\end{equation}
at $T_{C}=175$~MeV, including decays of $D^*$ mesons changes the
ratio to
\begin{equation}
{\frac{D^0}{D^+}} \approx \frac{1 + (1 + 0.68)\times 1.47}
{1 + 0.32 \times 1.47 }
 \approx 2.36,
\end{equation}
if we assume that the same branching ratios for $D^*$ decay to $D$
as in free space. The $D^0/D^+$ ratio in heavy-ion collisions is
thus about 25\% smaller than that in $pp$ collisions.

For the $\Lambda_c/D^0$ ratio, we have, by using $m_{\Lambda_c} =
2285$~MeV and $m_{D^0} = 1865$~MeV,
\begin{equation}
\left(\frac{\Lambda_c}{D^0}\right)_{0}
 = \frac{2 m_{\Lambda_c}^2 K_2(m_{\Lambda_c}/T_{C})}
{m_{D_0}^2 K_2(m_{D_0}/T_{C})} \simeq 0.24, \label{eq:thermal}
\end{equation}
where the factor $2$ is the spin degeneracy of $\Lambda_c$ baryon.
Since the ratio $D^*/D$ is 1.47, inclusion of $D^*$ resonance decays
causes a strong reduction of the $\Lambda_c/D^0$ ratio. On the other
hand, baryon resonances can also contribute to the production of
$\Lambda_c$ baryon through their decays and can enhance the
$\Lambda_c/D^0$ ratio. The major resonance contribution to
$\Lambda_c$ comes from $\Sigma_c^*(2520)$ of spin-$3/2$. In thermal
model, we have
\begin{equation}
\frac{\Sigma_c^*(2520)}{\Lambda_c} \approx 2 \times 3 \times
\frac{m_{\Sigma_c^*}^2
K_2(m_{\Sigma_c^*}^{}/T_{C})}{m_{\Lambda_c}^2
K_2(m_{\Lambda_c}^{}/T_{C})} = 1.8,
\end{equation}
where the factor $2$ comes from the difference in the spin
degeneracies of $\Lambda_c$ and $\Sigma_c^*$, and the factor $3$ is
due to the isospin $1$ of $\Sigma_c^*$. Since $\Sigma_c^*(2520)$
decays almost $100$\% into $\Lambda_c$, inclusion of its
contribution gives 180\% enhancement to the number of $\Lambda_c$.
Another important contribution to $\Lambda_c$ production is from the
decay of $\Sigma_c$ ground state,
$\Sigma_c(2455)$ of spin-1/2, via the $\Sigma_c \to \Lambda_c \pi$ decay.%
\footnote{In the strange quark sector, the small mass difference
between $\Lambda$ and $\Sigma$ does not allow the strong decay of
$\Sigma$ into $\Lambda$, and only $\Sigma^0 \to \Lambda\gamma$ is
allowed. But in heavy quark sector, the mass difference between
$\Lambda_c$ and $\Sigma_c$ is well above the pion mass.} In thermal
model, its abundance relative to that of $\Lambda_c$ is
\begin{equation}
\frac{\Sigma_c(2455)}{\Lambda_c} \approx 3 \times
\frac{m_{\Sigma_c}^2 K_2(m_{\Sigma_c}^{}/T_{C})}{m_{\Lambda_c}^2
K_2(m_{\Lambda_c}^{}/T_{C})} = 1.26.
\end{equation}
Therefore, the contribution from $\Sigma_c(2455)$ and
$\Sigma_c^*(2520)$ decays increases the $\Lambda_c$ number by a
factor of $3.1$ and brings the $\Lambda_c$ to $D^0$ ratio close to
the naive expectation given in Eq.~(\ref{eq:thermal}) based only on
directly produced $\Lambda_c$ and $D^0$, i.e.,
\begin{eqnarray}
\frac{\Lambda_c}{D^0} &=& \frac{\Lambda_c \{ 1 +
\Sigma_c^*(2520)/\Lambda_c + \Sigma_c(2455)/\Lambda_c\} }{D^0 (1 + 1.68 D^*/D)}
\Biggr|_{\rm ther}
\nonumber \\
&\simeq& \left(\frac{\Lambda_c}{D^0}\right)_{0}
\frac{ 1 + 1.8 + 1.26} {1+ 1.68 \times 1.47} \approx 0.28.
\label{eq:thermal2}
\end{eqnarray}

The contribution from higher mass charmed resonances is negligibly
small in the thermal model. For $D$ meson production, the next
higher state is $D_1(2420)$ of $J^P = 1^+$. The branching ratio of
the decay of this meson is not well-known except that it decays into
$D^* \pi$ but not into $D\pi$. If we assume that it decays into
$D^0$ via an intermediate $D^*$, i.e., $D_1 \to D^* \pi \to D
\pi\pi$, the denominator of Eq.~(\ref{eq:thermal2}) is then modified
by the addition of $1.68 \times 3 \times 0.06 \simeq 0.3$, where the
factor $0.06$ is the $D_1(2420)/D^0$ ratio, giving thus less than
10\% enhancement to $D^0$ production. For higher baryon resonances,
the next excited state is $\Lambda_c(2593)$. Its contribution to the
numerator of Eq.~(\ref{eq:thermal2}) is about $0.2$, since there is
no difference in its spin and isospin degeneracy from that of
$\Lambda_c$. Thus it again gives an effect less than 5\% to
$\Lambda_c$ production. The final value of the $\Lambda_c/D^0$ ratio
then essentially does not change, i.e., the ratio is modified from
$0.28$ to $0.27$, which is about twice of that in $pp$ collisions.

The estimated value of the $\Lambda_c/D^0$ ratio from our simple
thermal model is close to the value obtained in the more
sophisticated thermal or statistical model of Ref.~\cite{KR06b},
which gives a charmed baryon to meson ratio $(cqq)/(c\bar q) \le
0.25$. It is also consistent with the statistical model predictions
of Ref.~\cite{ABRS07}, which has $\Lambda_c/D^0 \sim 0.2$ at the
RHIC energy.

The $\Lambda_b/\bar{B}^0$ ratio in the thermal model can be
estimated in the same way as for the $\Lambda_c/D^0$ ratio. The main
difference between $\bar{B}^0$ and $D^0$ productions is that the
bottom quark is much heavier than the charm quark. As a result, the
mass difference between $B^*$ and $B$ is much smaller than that
between $D^*$ and $D$ because of the manifestation of heavy quark
spin symmetry. In fact, the mass difference between $D^*$ and $D$ is
about $140$~MeV, while that between $B^*$ and $B$ is
only~\cite{PDG08}
\begin{equation}
m(B^*) - m(B) \approx 5325 \mbox{ MeV} - 5279 \mbox{ MeV}
= 46 \mbox{ MeV}.
\end{equation}
Therefore, unlike $D^0$ production, $B^*$ mesons cannot decay into
$B$ mesons by strong interaction.%
\footnote{The decay of $B^*$ into $B$ is caused by electromagnetic
interactions, $B^* \to B\gamma$, and can thus be distinguished in
experiments.} This leads to a large suppression of $B$ meson
production compared with the case for $D$ meson production.
With $m(\Lambda_b) = 5620$~MeV, our simple thermal
model then gives
\begin{equation}
\frac{B^{*+}}{B^+} \approx 0.78, \qquad
\left(\frac{\Lambda_b}{\bar{B}^0} \right)_{0} \approx
0.32.
\end{equation}
It indeed shows that
$\left({\Lambda_b}/{\bar{B}^0} \right)_{0}$ is somewhat larger than
$\left({\Lambda_c}/D^0\right)_{0}$.

Unlike $B$ production, the production of $\Lambda_b$ is affected by
resonance decays. This is because the mass difference between
$\Lambda_b$ and $\Sigma_b$ approaches to a finite value ($\sim
195$~MeV~\cite{OPM94c-OP96a}) in the infinite mass limit, although
the mass difference between $\Sigma_b$ of spin-$1/2$ and
$\Sigma_b^*$ of spin-$3/2$ becomes smaller as the heavy quark mass
increases. By using the recent experimental information for the
$\Sigma_b$ baryons~\cite{CDF07},
\begin{equation}
M(\Sigma_b) = 5812~\mbox{ MeV}, \qquad
M(\Sigma_b^*) = 5833~\mbox{ MeV},
\end{equation}
the $\Lambda_b/\bar{B}^0$ ratio becomes
\begin{equation}
\frac{\Lambda_b}{\bar{B}^0} = \left( \frac{\Lambda_b}{\bar{B}^0}
\right)_{0} \left(1 + 3 \times 0.35+ 3 \times 2 \times 0.31
\right) \approx 1.25,
\end{equation}
where the factors 0.35 and 0.31 are the $\Sigma_b/\Lambda_b$ and
$\Sigma_b^*/\Lambda_b$ ratios, respectively. The
$\Lambda_b/\bar{B}^0$ ratio is seen to be larger than the
$\Lambda_c/D^0$ ratio by a factor of about $5$. Compared to its
value in $pp$ collisions, the $\Lambda_b/\bar{B}^0$ ratio in thermal
model for heavy ion collisions is close to a factor of two
larger, similar to the case for the $\Lambda_c/D^0$ ratio.

There are other higher charmed and bottom resonances listed by the
PDG~\cite{PDG08}. These include heavy baryons that contain strange
quarks such as $\Xi_Q$, $\Xi_Q'$, $\Omega_Q$, and their resonances.
Although most of corresponding bottom baryons are yet to be
discovered, their existence has been predicted by the quark
model~\cite{KKLR08}. Using the masses listed by the PDG and from the
quark model, we find that the yield of these baryons is about 60\%
of the $\Lambda_Q$ yield in the thermal model and is thus
appreciable. Including these higher resonances in the thermal model
changes the baryon/meson ratios to
\begin{equation}
\frac{\Lambda_c}{D^0} \simeq 0.27, \qquad \frac{\Lambda_b}{\bar{B}^0} \simeq
0.86.
\end{equation}

\section{The coalescence model}
\label{sec:coal}

In the thermal model, the relative abundance of particles depends
only on their masses. Whether diquarks exist in QGP thus does not
affect the $\Lambda_c/D^0$ and $\Lambda_b/\bar{B}^0$ ratios in the
thermal model. This is not the case in the quark coalescence model
as to be shown in the rest of this paper.

\subsection{Formalism}

As in the coalescence model described in Refs.~\cite{GKL03,CK06},
which has been used in Ref.~\cite{GKR04} to study the $D$ meson
transverse momentum spectrum in relativistic heavy ion collisions,
we assume that quarks, antiquarks, and gluons in the produced QGP
are uniformly distributed in a fire cylinder of volume $V$. Their
momentum distributions are taken to be thermal in the transverse
direction but uniform in midrapidity. For
the Wigner functions of hadrons, they are expressed in terms of
Gaussian functions. Modeling the effect of transverse radial flow of
the fire cylinder by an effective temperature, the produced heavy
meson transverse momentum spectrum is then given by
\begin{eqnarray}
\frac{dN_M}{d\bm{p}_M^{}} &=& g_M^{} \frac{(2\sqrt{\pi}\sigma)^3}{V}
\int d \bm{p}_1^{} d \bm{p}_{2}^{} \frac{dN_1}{d\bm{p}_1^{}}
\frac{dN_{2}}{d\bm{p}_{2}^{}} \nonumber \\ && \mbox{} \times
\exp\left(-\bm{k}^2 \sigma^2 \right) \delta( \bm{p}_M^{} -
\bm{p}_1^{} - \bm{p}_{2}^{}), \label{eq:meson-coal}
\end{eqnarray}
where $g_M^{}$ is the statistical factor for colored quark and
antiquark to form a color neutral meson, e.g., $g_{D^0}^{}=1/36$ and
$g_{D^{*0}}^{}=1/12$ for $D^0$ and $D^{*0}$, respectively, and
similarly for ${\bar B}^0$ and ${\bar B}^{*0}$. The distribution of
heavy quarks with transverse momentum $\bm{p}_1$ and light
antiquarks with transverse momentum $\bm{p}_2$ in the fire cylinder
frame are denoted by $dN_1/d\bm{p}_1$ and $dN_2/d\bm{p}_2$,
respectively. The $\bm{p}_M^{}$ is the transverse momentum of
produced heavy meson. The relative transverse momentum $\bm{k}$
between the heavy quark and light antiquark is defined as
\begin{equation}
\bm{k} = \frac{1}{m_1^{} + m_2^{}}
\left( m_2^{} \bm{p}_1' - m_1^{} \bm{p}_2' \right),
\label{eq:def_k}
\end{equation}
where $m_{1,2}^{}$ are quark masses, and $\bm{p}_1'$ and $\bm{p}_2'$
are heavy quark and light antiquark transverse momenta,
respectively, defined in the center-of-mass frame of produced heavy
meson. The width parameter $\sigma$ is related to the harmonic
oscillator frequency $\omega$ by $\sigma = 1/\sqrt{\mu\omega}$,
where $\mu = m_1^{} m_2^{} / (m_1^{} + m_2^{})$ is the reduced mass
and can be related to the size of produced hadrons. Namely, the
charge rms radius of produced meson is given by
\begin{equation}
\langle r^2 \rangle_{\rm ch} = \frac{3}{2} \frac{1}{\mu\omega}
\frac{Q_1^{} m_2^2 + Q_2^{} m_1^2}{(m_1^{} + m_2^{})^2},
\end{equation}
with $Q_i^{}$ being the charge of the $i$-th (anti)quark. To
reproduce the root-mean-squared charge radii $0.43$~fm of $D^+$ and
$0.62$~fm of $B^+$ as predicted by the light-front quark model of
Ref.~\cite{Hwang02}, an oscillator frequency $\omega=0.33$~GeV is
needed.

We note that contrary to that for mesons consisting of only light
quarks, the coalescence formula, Eq.~(\ref{eq:meson-coal}), does not
require heavy and light quarks to have similar momenta to form heavy
hadrons. Instead, quarks of similar velocities have the most
probable chance to form hadrons.

Similarly, the coalescence formula for forming $\Lambda_Q$,
$\Sigma_Q$, $\Xi_Q$, $\Xi_Q'$, and $\Omega_Q$ as well as their
resonances from three quarks are given by
\begin{eqnarray}
\frac{dN_B}{d\bm{p}_{B}^{}} &=& g_{B}^{} \frac{(2\sqrt{\pi})^6
(\sigma_1^{} \sigma_2^{})^3}{V^2} \int d \bm{p}_1^{} d \bm{p}_2^{} d
\bm{p}_3 \frac{dN_1}{d\bm{p}_1^{}} \frac{dN_2}{d\bm{p}_2^{}}
\frac{dN_3}{d\bm{p}_3^{}} \nonumber \\ && \mbox{} \times \exp\left(
-\bm{k}_1^2 \sigma_1^2 - \bm{k}_2^2 \sigma_2^2 \right) \delta(
\bm{p}_{B}^{} - \bm{p}_1^{} - \bm{p}_2^{} - \bm{p}_3^{}),
\nonumber \\
\label{eq:3q-coal}
\end{eqnarray}
where the index $3$ refers to the heavy quark and indices $1$ and
$2$ to light quarks, and the statistical factor $g_B^{}$ has values
$1/108$, $1/36$, and $1/18$ for $\Lambda_Q$, $\Sigma_Q$, and
$\Sigma_Q^*$, respectively, and $1/54$, $1/18$, $1/108$, and $1/36$
for $\Xi_Q$ ($\Xi_Q')$, $\Xi_Q^*$, $\Omega_Q$, and $\Omega_Q^*$,
respectively. The relative transverse momenta defined in the
center of mass frame of produced baryon are
\begin{eqnarray}
\bm{k}_1 &=& \frac{1}{m_1^{} + m_2^{}}
\left( m_2^{} \bm{p}_1' - m_1^{} \bm{p}_2' \right),
\nonumber \\
\bm{k}_2 &=& \frac{1}{m_1^{} + m_2^{} + m_3^{}}
\left[ m_3^{} \left( \bm{p}_1' + \bm{p}_2' \right)
- (m_1^{} + m_2^{}) \bm{p}_3' \right].
\nonumber \\
\end{eqnarray}
The width parameters $\sigma_i^{}$ are $\sigma_i^{} =
1/\sqrt{\mu_i\omega}$, where
\begin{eqnarray}
\mu_1^{} = \frac{m_1^{} m_2^{}}{m_1^{} + m_2^{}}, \qquad
\mu_2^{} = \frac{(m_1^{} + m_2^{}) m_3^{}}{m_1^{} + m_2^{} +
m_3^{}},
\end{eqnarray}
and the oscillator frequency $\omega$ is related to the baryon
charge radius by
\begin{eqnarray}
\langle r^2 \rangle_{\rm ch} &=& \frac{3}{2\omega} \frac{1}{m_1^{} +
m_2^{} + m_3^{}} \left( \frac{m_2^{} + m_3^{}}{m_1^{}} Q_1
\right.\nonumber \\ && \left. \quad \mbox{} + \frac{m_3^{} +
m_1^{}}{m_2^{}} Q_2 + \frac{m_1^{} + m_2^{}}{m_3^{}} Q_3 \right).
\end{eqnarray}
The root-mean-squared charge radii of $\Lambda_c$ and
$\Lambda_b$ predicted by the quark model have similar values of
about 0.39 fm~\cite{brac96}, and they are reproduced by the
oscillator frequencies $0.43$~GeV and $0.41$~GeV, respectively. These
values are 25-30\% larger than the ones for heavy mesons. As we
shall show in the next subsection, to convert as many heavy quarks
of small $p_T^{}$
into heavy hadrons via coalescence as possible%
\footnote{This is similar to the assumption introduced in the
thermal model that all charmed quarks are converted into heavy
hadrons.}
requires, however, smaller oscillator frequencies or
larger heavy hadron radii.

For production of $\Lambda_Q$ baryons from the coalescence of heavy
quarks and $[ud]$ diquarks, we consider $\Lambda_Q$ baryons as made
of a heavy quark and a light diquark only. Their spectrum can thus
be obtained from Eq.~(\ref{eq:meson-coal}) by replacing $g_M^{}$
with $g_B' = 1/9$. Explicitly, it is written as
\begin{eqnarray}
\frac{dN_B}{d\bm{p}_B^{}} &=& g_B' \frac{(2\sqrt{\pi}\sigma_{dq}^{})^3}{V}
\int d \bm{p}_1^{} d \bm{p}_{2}^{}
\frac{dN_1}{d\bm{p}_1^{}}
\frac{dN_{2}}{d\bm{p}_{2}^{}}
\nonumber \\ && \mbox{} \times
\exp\left(-\bm{k}^2 \sigma_{dq}^2 \right)
\delta( \bm{p}_B^{} - \bm{p}_1^{} - \bm{p}_{2}^{}).
\label{eq:dq-coal}
\end{eqnarray}
The width parameter $\sigma_{dq}^{}$ or corresponding oscillator
frequency $\omega_{dq}$ is determined by fitting the sizes of heavy
baryons used in the three-quark model.

The scalar diquarks can also lead to formation of excited states of
$\Lambda_Q$ baryons with $j^\pi = \frac12^-$ and $\frac32^-$, which
can decay into $\Lambda_Q$ by emitting pions. We note that the mass
differences between $j^\pi = \frac12^-$ and $\frac32^-$ states are
small because of heavy quark spin symmetry, e.g., the lowest excited
states of $\Lambda_c$ are $\Lambda(2593)$ of $j^\pi = \frac12^-$ and
$\Lambda_c(2625)$ of $j^\pi = \frac32^-$, and thus have a mass
difference of only $32$~MeV. Although the number of these resonances
is rather small in the thermal model due to their higher masses
compared to those of ground state $\Lambda_Q$ baryons, it is not the
case in the coalescence model if diquarks exist in QGP. Since
excited $\Lambda_Q$ baryons have negative parities, the scalar
diquark inside them is in a relative $P$-wave state with respect to
the heavy quark. The coalescence formula for the production of such
hadrons is
\begin{eqnarray}
\frac{dN_B}{d\bm{p}_B^{}} &=& g_B' \frac{2(2\sqrt{\pi}\sigma_{dq}^{})^3}{3V}
\int d \bm{p}_1^{} d \bm{p}_{2}^{}
\frac{dN_1}{d\bm{p}_1^{}}
\frac{dN_{2}}{d\bm{p}_{2}^{}}
\sigma_{dq}^2 k^2
\nonumber \\ && \mbox{} \times
\exp\left(-\bm{k}^2 \sigma_{dq}^2 \right)
\delta( \bm{p}_B^{} - \bm{p}_1^{} - \bm{p}_{2}^{}),
\label{eq:dq-coal2}
\end{eqnarray}
using the Wigner function given in Refs.~\cite{BD96a,CKLN07}. In the
above, $k$ is similarly defined as in Eq.~(\ref{eq:def_k}) and the
width parameter $\sigma_{dq}^{}$ is taken to have same value as that
for the ground state $\Lambda_Q$.

\subsection{Quark distribution functions in QGP}

For the quark distribution functions in quark-gluon plasma, we use
the thermal distribution for light quarks, namely,
\begin{equation}
\frac{dN_q}{d\bm{p}} = \lambda_q^{} \frac{g_q^{} V}{(2\pi)^3} m_T^{}
\exp(-m_T^{}/T_{\rm eff}), \label{eq:quark_dist}
\end{equation}
where $g_q^{} = 6$, $m_T^{} = \sqrt{\bm{p}^2 + m_q^2}$, and $\bm{p}$
is the transverse momentum. Instead of the local temperature and
transverse flow velocity, we use an effective temperature $T_{\rm
eff}$ ($= 200$~MeV) and introduce a normalization factor
$\lambda_q^{}$ to fix the total number of quarks as in
Ref.~\cite{OK07}. Following Ref.~\cite{CGKL03}, we assume that the
phase transition temperature is $T_C^{}= 175$~MeV and the volume of
the quark-gluon plasma formed in rapidity $|y|\le 0.5$ is
$V=1,000~{\rm fm}^3$ for central Au$+$Au collisions at
$\sqrt{s_{NN}^{}}=200$~GeV at RHIC. This gives the light
quark numbers $N_u = N_d \approx 245$ and the strange quark number
$N_s\approx 149$ if the constituent light quark and strange quark
masses are taken to be 300 MeV and 475 MeV, respectively. The
resulting normalization factors $\lambda_q$ for light and strange
quarks are then $0.95$ and $0.81$, respectively.

Diquarks are also assumed to be in thermal equilibrium and their
distribution $dN_{[ud]}/d\bm{p}^{}$ is then the same as given in
Eq.~(\ref{eq:quark_dist}) with $g_{[ud]}^{} = 3$. The number of
diquarks depends on the diquark mass and is in the range of $77 \sim
44$ with $m_{[ud]}^{} = 455 \sim 600$~MeV~\cite{LOYYK07}. If
diquarks exist in QGP, the number of free light quarks is reduced by
twice the number of diquarks so that the total number of light
quarks is preserved. This would cause a reduction of meson yields
and an enhancement of baryon yields and thus induces larger
$\Lambda_Q/H^0$ ratios. Results to be presented below are obtained
by using the diquark mass $m_{[ud]}^{} = 455$~MeV, which is in the
range of diquark mass estimated in Ref.~\cite{CRP87}. The
sensitivity of these results to the diquark mass will, however, be
briefly discussed.

For heavy quark transverse momentum distributions in midrapidity, we
adopt those parameterized in Ref.~\cite{KL07} for central Au$+$Au
collisions at $\sqrt{s_{NN}^{}}=200$~GeV at RHIC, i.e.,
\begin{eqnarray}
\frac{dN_c}{d\bm{p}} &=& \frac{19.2 \left[ 1 + (p/6)^2 \right]} {(1 +
p/3.7)^{12} \left[ 1 + \exp(0.9 - 2p) \right]},
\nonumber \\
\frac{dN_b}{d\bm{p}} &=& 0.0025 \left[ 1 +
\left(\frac{p}{16}\right)^5 \right] \exp(-p/1.495),
\end{eqnarray}
with the transverse momentum $p$ in unit of GeV. These distributions
were obtained from heavy quark $p_T^{}$ spectra from $pp$ collisions
at same energy by the number of binary collisions ($\sim 960$) in
Au+Au collisions. For bottom quarks, it was taken from the upper
limit of the uncertainty band of the pQCD predictions in
Ref.~\cite{CNV05} for $pp$ collisions, as no experimental
measurements are available at RHIC. The charm quark $p_T^{}$
spectrum in $pp$ collisions is, on the other hand, determined from
fitting simultaneously measured $p_T^{}$ spectrum of charmed mesons
from d+Au collisions~\cite{STAR04d} and of electrons from heavy
meson decays in $pp$ collisions.

Including heavy quark energy loss as estimated in Ref.~\cite{KL07},
the resulting heavy quark transverse momentum distribution in
midrapidity can be parameterized as
\begin{equation}
\frac{dN_Q^{\rm EL}}{d\bm{p}} = \frac{dN_Q}{d\bm{p}} L_Q(p),
\end{equation}
with
\begin{eqnarray}
L_c &=& 0.8 \exp(-p/1.2) + 0.6 \exp(-p/15),
\nonumber \\
L_b &=& 0.36 + 0.84 \exp(-p/10),
\label{eq:loss}
\end{eqnarray}
for charm and bottom quarks, respectively, and $p$ again in GeV. The
heavy quark numbers are then estimated to be $N_c \simeq 9.23$ and
$N_b \simeq 0.035$. The charm quark number used here is about a
factor three larger than that used in Ref.~\cite{LOYYK07} based on
perturbative QCD calculations. This, however, does not affect the
$\Lambda_Q/H^0$ ratios. For the heavy quark
masses, we use $m_c^{} = 1.5$~GeV and $m_b^{} = 5.0$~GeV.

\subsection{Results}

\begin{table*}[t]
\centering
\begin{tabular}{c|ccccc|cccccc}  \hline \hline
  & \multicolumn{5}{c|}{Charmed hadrons} &
  \multicolumn{6}{c}{Bottom hadrons } \\ \hline
 Model & $D^0$ & $D^+$ & $D_s$ & $\Lambda_c$ & $\Xi_c/\Xi_c'/\Omega_c$
 & $\bar{B}^0$ ($B^-$) & $\bar{B}^{*0}$ ($B^{*-}$) &
 $B_s^0$ & $B_s^{*0}$ & $\Lambda_b$ & $\Xi_b/\Xi_b'/\Omega_c$ \\
\hline
PYTHIA model & 0.607 & 0.196 & 0.121 & 0.076 &  &
  0.101 & 0.3015 & 0.030 & 0.089 & 0.076 \\ \hline
Thermal model & 0.435 & 0.205 & 0.179 & 0.118 & 0.063 &
  0.111 & 0.229 & 0.049 & 0.113 & 0.096 & 0.062 \\ \hline
Coalescence model (three-quark model) & 0.348 & 0.113 & 0.113 & 0.288 & 0.138
 & 0.053 & 0.158 & 0.027 & 0.080 & 0.316 & 0.155 \\
  (ground state contribution) & 0.051 & 0.051 & 0.0264 & 0.028 & 0.066
  & 0.052 & 0.155 & 0.026 & 0.079 & 0.032 & 0.073 \\
  (fragmentation) & 0.043 & 0.014 & 0.009 & 0.005 &
  & 0.001 & 0.003 & 0.0003 & 0.001 & 0.001 \\ \hline
Coalescence model (diquark model) & 0.282 & 0.091 & 0.123 & 0.378 & 0.126
 & 0.044 & 0.131 & 0.031 & 0.092 & 0.385 & 0.142 \\
 (ground state contribution) & 0.039 & 0.039 & 0.028 & 0.016 & 0.059
 & 0.040 & 0.120 & 0.029 & 0.088 & 0.019 & 0.066 \\
 (diquark contribution) & & & & 0.205 &
 & & & & & 0.192 & \\
 (fragmentation) & 0.048 & 0.015 & 0.010 & 0.005 &
 & 0.004 & 0.011 & 0.002 & 0.004 & 0.003 & \\
 \hline \hline
\end{tabular}
\caption{\label{tab:ratio} Relative production fractions of
midrapidity heavy hadrons produced in central Au+Au collisions at
$\sqrt{s_{NN}}=200$ GeV. The heavy-strange baryon ($\Xi_Q$,
$\Xi_Q'$, and $\Omega_Q$) yield is suppressed in the PYTHIA model
and thus neglected in this work.}
\end{table*}

To calculate the spectra of $H^0$ and $\Lambda_Q$ produced in
relativistic heavy ion collisions, we consider two scenarios for
$\Lambda_Q$ production. In the first model, we assume that the
$\Lambda_Q$ baryons have no diquark structure and there are no
diquarks in QGP either. Therefore, $\Lambda_Q$, $\Sigma_Q$,
$\Sigma_Q^*$, $\Xi_Q$, $\Xi_Q'$, $\Xi_Q^*$, $\Omega_Q$, and
$\Omega_Q^*$ baryons are all formed by three-quark coalescence.
We further include heavy mesons that consist of strange quarks 
such as $D_s$
and $B_s$ and their resonances. Using the same oscillator
frequency for both heavy mesons and baryons as shown approximately
in the quark model, we adjust its value to convert all heavy
quarks of small $p_T^{}$ into heavy hadrons via coalescence as
in the thermal model. This leads to $\omega =
0.106$~GeV for charmed hadrons and $\omega = 0.059$~GeV for bottom
hadrons. The resulting $D^+$ and $B^+$ charge radii are $0.74$~fm
and $1.44$~fm, respectively, which are factors of $1.7$ and $2.3$
larger than those predicted by
the light-front quark model of Ref.~\cite{Hwang02}. 
The required smaller oscillator frequencies may be partly due
to the change in hadron radii in medium~\cite{LTTW97b} and
partly due to the incomplete treatment of production and decay of
hadron resonances in the present study.

The second model assumes the diquark structure in $\Lambda_Q$ and
the existence of diquarks in QGP as in Ref.~\cite{LOYYK07}. In this
case, $\Lambda_Q$ baryons are formed by diquark--heavy-quark coalescence
in addition to three-quark coalescence.%
\footnote{Strictly speaking, the three quarks should coalesce to
form the $\Lambda_Q$ in diquark structure. Here we simply
approximate this production process by three-quark coalescence to
form non-diquark $\Lambda_Q$. This thus gives an upper bound for the
yield of $\Lambda_c$ from three independent quarks in QGP as pointed
out in Ref.~\cite{LOYYK07}.} Together with the $\Sigma_Q$ and
$\Sigma_Q^*$ resonance contributions as well as those from $\Xi_Q$,
$\Xi_Q'$, $\Xi_Q^*$, $\Omega_Q$, and $\Omega_Q^*$ resonances, which
are formed only from three-quark coalescence as they do not contain
the scalar diquark structure, excited $\Lambda_Q(\frac12^-)$ and
$\Lambda_Q(\frac32^-)$ baryons are also included using the $P$-wave
diquark--heavy-quark
coalescence.%
\footnote{In general, coalescence of three independent quarks can
also form negative parity baryons. Since this process has a smaller
contribution than that due to the production of positive parity
baryons~\cite{KM06b}, it is not considered in present work.} As in
the three-quark coalescence model, the oscillator frequencies or
width parameters are determined by requiring that all low $p_T^{}$
heavy quarks are converted to hadrons. Besides the common oscillator
frequency $\omega$ for mesons and three-quark baryons, the
additional oscillator frequency for the heavy-quark and diquark
system is fixed by reproducing the same $\Lambda_Q$ radius used in
the three-quark configuration. Then we have $\omega = 0.09$~GeV for
charmed hadrons and $\omega = 0.049$~GeV for bottom hadrons, leading
to $0.81$~fm and $1.58$~fm for the $D^+$ and $B^+$ charged radii,
respectively. These values lead to $\omega_{dq}^{} = 0.053$~GeV and
$0.029$~GeV, respectively, for charmed and bottom baryons consisting
of diquarks. In obtaining the contribution of resonances to the
ground state $H^0$ and $\Lambda_Q$ spectra, we neglect for
simplicity the small momentum shift in their decays.

In both models, remaining heavy quarks that are not converted to
hadrons via coalescence, which mostly have high $p_T^{}$, are converted
to heavy hadrons by fragmentation. This is achieved by assuming that
a heavy quark of transverse momentum $p_T^{}$ fragments into
$\Lambda_Q$ and $H^0$ with a ratio similar to that given by the
PYTHIA model, i.e.,
\begin{eqnarray}
\left( \frac{\Lambda_c}{D^0} \right)_{\rm fr} &=&
0.05 \exp\left[-(p_T^{}-4.0)^2/8.0 \right] + 0.1,
\nonumber \\
\left( \frac{\Lambda_b}{\bar{B}^0} \right)_{\rm fr} &=&
0.75,
\label{eq:param}
\end{eqnarray}
with $p_T^{}$ in unit of GeV.

\subsubsection{Relative production fractions of heavy hadrons
and $\Lambda_Q/H^0$ ratios}

Our results for the relative production fractions of heavy hadrons
produced in central Au+Au collisions at $\sqrt{s_{NN}^{}}=200$~GeV
are presented in Table~\ref{tab:ratio} together with those from the
PYTHIA model and the thermal model. It is seen that the relative
production fraction of $D^0$ among all produced charmed hadrons in
the three-quark coalescence model is $0.35$, while it is $0.44$ and
$0.61$ for the thermal model and the PYTHIA model, respectively.
This fraction reduces to $0.28$ in the diquark model. The relative
production fraction of $\bar{B}^0$ among all produced bottom hadrons
also shows a similar behavior, namely, it is $0.101$, $0.097$,
$0.052$, and $0.043$ in the PYTHIA model, the thermal model, the
three-quark model, and the diquark model, respectively.

For the $\Lambda_Q/H^0$ ratios, we find that without resonance
contribution $\Lambda_c/D^0 = 0.55$ in the three-quark model and
$\Lambda_c/D^0 = 2.2$ in the diquark model. These results are
different from those of Ref.~\cite{LOYYK07}, namely, the
$\Lambda_c/D^0$ is larger and the enhancement of this ratio due to
the presence of diquarks is about $4$ while it was about $8$ in
Ref.~\cite{LOYYK07}. These differences come from the different
values for the oscillator frequencies or width parameters as well as
the fact that in Ref.~\cite{LOYYK07} the number of free quarks in
the quark-gluon plasma is not reduced by the presence of diquarks.
For bottom hadrons, we obtain $\Lambda_{b}/\bar{B}^0 = 0.6$ in the
three-quark model, which increases by a factor of $3.5$ to
$\Lambda_{b}/\bar{B}^0 = 2.1$ in the diquark model.

Including resonances enhances both $\Lambda_Q$ and $H^0$ production.
We find that in the three-quark model, about 70\% of $D^0$ and about
90\% of $\Lambda_Q$ are produced through resonance decays. The
importance of resonances is also seen in the diquark model. In
particular, the diquark contribution through negative parity
$\Lambda_Q$ resonances is non-negligible. The number of negative
parity $\Lambda_Q$ resonance of spin-$1/2$ turns out to be about
half of the ground state $\Lambda_Q$ baryon number. Since
$\Lambda_Q(\frac12^-)$ and $\Lambda_Q(\frac32^-)$ resonances are
almost degenerate in mass, the number of negative parity $\Lambda_Q$
resonances is about two times that of the ground state $\Lambda_Q$
after taking into account the spin degeneracy. This should be
contrasted with the thermal model which predicts suppressed yields
of negative parity $\Lambda_Q$ resonances as a result of their
larger masses. As a result, we find that about 70\% of $D^0$ and
about 75\% of $\Lambda_Q$ are produced through resonance decays in
the diquark model.

The contribution from heavy quark fragmentation is small,
particularly to the ratios of particle numbers. This is because we
have assumed that most hadrons are produced by coalescence of low
$p_T^{}$ heavy quarks. The fragmentation contribution is thus
non-negligible only for high $p_T^{}$ hadrons, which constitute less
than 8\% of produced total heavy hadrons.

The final $\Lambda_Q/H^0$ ratios after including fragmentation and
resonance contributions are
\begin{equation}
\Lambda_c/D^0 = 0.83, \qquad \Lambda_b/\bar{B}^0 = 6.00,
\end{equation}
for the three-quark model, and
\begin{equation}
\Lambda_c/D^0 = 1.34, \qquad \Lambda_b/\bar{B}^0 = 8.79,
\end{equation}
for the diquark model. Therefore, the enhancement factor for these
ratios due to diquarks in QGP is about $1.6$ for both charmed and
bottom hadrons.

Our results thus indicate that the $\Lambda_c/D^0$
($\Lambda_b/\bar{B}^0$) ratio in heavy ion collisions predicted by
the coalescence model is about a factor of 6.4 (8.6) larger than the
predictions of the PYTHIA model in Section~\ref{sec:pp} and about a
factor of 3.1 (7.0) larger than that from the thermal model in
Section~\ref{sec:thermal}. The existence of diquarks in QGP further
enhances this factor to 10.3 (12.6) and 5.0 (10.2), compared to
those from the PYTHIA and thermal models, respectively.

\subsubsection{Transverse momentum spectra of heavy hadrons}

\begin{figure}[t]
\centering
\includegraphics[width=\columnwidth, angle=0,clip]{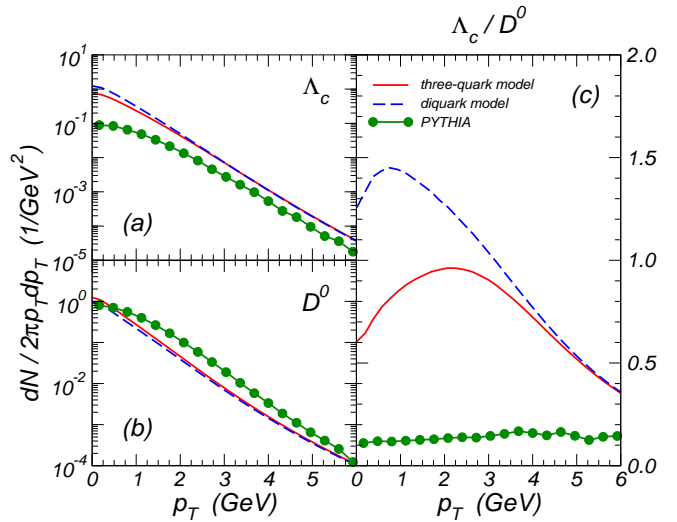}
\caption{\label{fig:charm-fr} Spectra of (a) $\Lambda_c$ and (b)
$D^0$, and (c) the ratio $\Lambda_c/D^0$ in midrapidity
($|y|\le 0.5$) for central Au+Au collisions at
$\sqrt{s_{NN}^{}}=200$ GeV. Solid lines are for the three-quark
model and dashed lines are for the diquark model. Results from the
PYTHIA model are shown by filled circles.}
\end{figure}

\begin{figure}[t]
\centering
\includegraphics[width=\columnwidth, angle=0,clip]{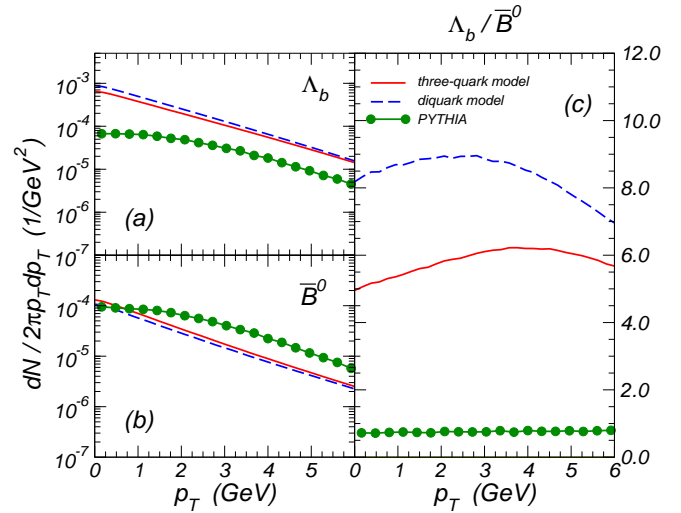}
\caption{\label{fig:bottom-fr} Same as Fig.~\ref{fig:charm-fr} for
(a) $\Lambda_b$ spectrum, (b) $\bar{B}^0$ spectrum, and (c) the
$\Lambda_b/\bar{B}^0$ ratio.}
\end{figure}

The transverse momentum spectra of $\Lambda_Q$ and $H^0$ are shown
in Figs.~\ref{fig:charm-fr} and \ref{fig:bottom-fr} for charmed and
bottom hadrons, respectively, together with the transverse momentum
dependence of the $\Lambda_Q/H^0$ ratio. Solid lines are results for
the three-quark model and dashed lines are those for the diquark
model. Compared with the PYTHIA results shown by filled circles, the
enhancement of $\Lambda_Q$ yield and the suppression of $H^0$ yield
are evident. The bottom hadron spectra are seen to be harder than
the charmed hadron spectra due to the flatter bottom quark spectrum
than the charm quark spectrum.

Compared with measured $\bar{p}/\pi^-$ and $\Lambda/K_S^0$
ratios~\cite{PHENIX04b,STAR06d,Soren06}, the obtained
$\Lambda_Q/H^0$ ratios show very different behavior. Although the
$\Lambda_c/D^0$ ratio has a similar shape as the $\bar{p}/\pi^-$ and
$\Lambda/K_S^0$ ratios, which could be approximated by a Gaussian
shape as assumed in Ref.~\cite{MGC07}, it has a much larger width
and is thus much flatter than the light baryon/meson ratios.
Furthermore, the $\Lambda_b/\bar{B}^0$ ratio is even flatter than
the $\Lambda_c/D^0$ ratio. This behavior can also be found in the
spectra of $\Lambda_Q$ and $H^0$. Namely, while the spectra of
$\Lambda_b$ and $\bar{B}^0$ have very similar $p_T^{}$ dependence
(Fig.~\ref{fig:bottom-fr}), those of $\Lambda_c$ and $D^0$ show
somewhat different $p_T^{}$ dependence (Fig.~\ref{fig:charm-fr}).
All these behaviors originate from the large mass of heavy quarks.
In the infinite mass limit, the momentum of a heavy hadron is
completely carried by the heavy quark, and the role of light quarks
or diquarks is just to give different production probability
depending on the heavy hadron wavefunction. Therefore, the
$\Lambda_Q/H^0$ ratio in the infinite heavy mass limit becomes a
constant and shows no momentum dependence. In the case of the
$\Lambda_c/D^0$ ratio, finite mass effects are not negligible but
its dependence on the transverse momentum is less strong than in the
$\Lambda/K_S^0$ ratio. This is clearly seen in $pp$ collisions as
shown by the PYTHIA model results in Figs.~\ref{fig:charm-fr} and
\ref{fig:bottom-fr}. Although the coalescence model shows a $p_T^{}$
dependence of the $\Lambda_Q/H^0$ ratio, this dependence weakens as
the heavy quark mass increases.

We also find that inclusion of diquarks in QGP causes a shift of the
peak positions in the $\Lambda_Q/H^0$ ratios to lower values of
$p_T^{}$. In the three-quark model, the $\Lambda_c/D^0$ ratio peaks
at $p_T^{} \simeq 2$~GeV, which shifts to $p_T^{} \simeq 0.8$~GeV in
the diquark model.  The change of the peak position is caused by
both the reduction of the $D^0$ meson spectrum and the enhancement
of the $\Lambda_c$ spectrum in the diquark model. We find, however,
that the latter has a larger effect on the peak position in the
$\Lambda_c/D^0$ ratio. Since the enhancement of $\Lambda_c$ due to
light diquarks is more appreciable in the low $p_T^{}$ region, the
peak of the $\Lambda_c/D^0$ ratio in the diquark model thus appears
at a lower value of $p_T^{}$ compared to that in the three-quark
model. The $\Lambda_b/\bar{B}^0$ ratio also shows a shift in the
peak position, i.e., from $p_T^{} \simeq 3.5$~GeV in the three-quark
model to $2$~GeV in the diquark model, but it is still much flatter
than the $\Lambda_c/D^0$ ratio.

For a more realistic momentum distribution of heavy hadrons in
relativistic heavy ion collisions after their production from QGP,
their production and annihilation due to collisions of and with
surrounding particles should be investigated by a transport model
like the {\sc ampt} model~\cite{ZKLL00a,LKLZP04}. However, cross
sections for reactions involving open charmed hadrons are
small~\cite{LK02,lin}, and it has been shown in Ref.~\cite{CKLN07}
that the effects of multiple scattering would be small with such
small cross sections. Since this work is at the level of an
exploratory study, such an elaborated investigation is not yet
called for.

\section{Conclusions and Discussions}
\label{sec:summary}

In this paper, we have studied the $\Lambda_Q/H^0$ ($\Lambda_c/D^0$
and $\Lambda_b/\bar{B}^0$) ratios in heavy ion collisions,
particularly the enhancement of these ratios due to coalescence of
heavy quarks with light quarks as well as diquarks that might exist
in the produced QGP. We have also considered the resonance decay
effects which are shown to be important for understanding the
particle production ratios in $pp$ collisions. Our simple estimate
based on the thermal model as well as predictions of more
sophisticated statistical models~\cite{KR06b,ABRS07} show that
resonance decays also plays a crucial role for heavy hadron
production in heavy ion collisions. We have, therefore, included the
contribution from resonances in estimating these ratios.

We have determined the width parameters of hadron Wigner functions
used in the coalescence model by requiring that the majority of
heavy hadrons at low $p_T^{}$ are formed from coalescence, which is
similar to that assumed in the thermal model. The remaining heavy
quarks are then converted to heavy hadrons via fragmentation as in
$pp$ collisions. The resulting $\Lambda_c/D^0$ and
$\Lambda_b/\bar{B}^0$ ratios in central Au+Au collisions at
$\sqrt{s_{NN}^{}}=200$ GeV are found to be $1.34$ and $8.79$,
respectively, which are about $10.3$ and $12.6$ larger than the
predictions of the PYTHIA model, and about a factor of $5.0$ and
$10.2$ larger than those of the thermal model. The enhancement of
the $\Lambda_Q/H^0$ ratio due to diquarks is, however, only about
1.6 for both charmed and bottom hadrons, which is smaller than that
predicted in Ref.~\cite{LOYYK07}. The difference mainly arises from
the different constraints used here and in Ref.~\cite{LOYYK07} for
determining the oscillator frequencies or width parameters in the
hadron Wigner functions and from maintaining the same total light
quark numbers in both the three-quark model and diquark model in the
present study.

The enhancement due to diquarks would have been larger if we do not
take into account the resonance contribution. This is because of the
enhancement of $\Lambda_Q$ baryons from $\Sigma_Q$ and $\Sigma_Q^*$
decays, whose production is not affected by the presence of diquarks
in QGP. This is different from the production of strange $\Lambda$
to which $\Sigma$ can not decay by strong interactions. Furthermore,
including the resonance contribution increases the
$\Lambda_b/\bar{B}^0$ ratio much more than the $\Lambda_c/D^0$ ratio
because $B^*$ cannot decay into $B$ by strong interactions as a
result of heavy quark spin symmetry. Finally, this model leads to
the reduction of $D^0$ and $\bar{B}^0$ mesons by factors of about
$2.2$ compared to the PYTHIA model.

Contrary to the PYTHIA model, we have found that the yields of heavy
baryons with strange quark(s) are not small. Their yield was
estimated to be about 60\% of the $\Lambda_Q$ yield in the thermal
model. In the coalescence model, this is about 50\% and 35\% of the
$\Lambda_Q$ yield in the three-quark model and diquark model,
respectively.

We have also compared the transverse momentum dependence of the
$\Lambda_Q/H^0$ ratios with that of measured $\Lambda/K_S^0$ ratio
and found that the $\Lambda_Q/H^0$ ratios have a weaker dependence
on the transverse momentum because of the massive quarks inside
heavy hadrons. In particular, this leads to a very weak dependence
of the $\Lambda_b/\bar{B}^0$ ratio on the transverse momentum.

We have further found that the $\Lambda_Q/D^0$ ratio peaks at
$p_T^{} \simeq 2$~GeV in the three-quark model and at $p_T^{} \simeq
0.8$~GeV in the diquark model. Therefore, the enhancement of heavy
baryon production over heavy meson production due to diquarks can
mostly be observed at low $p_T^{}$ region. We have also estimated
that the ratio $\Lambda_c/\Sigma_c^0 \simeq 23$, which is less than
that of Ref.~\cite{Sateesh92} by a factor of $\sim 4$.

These results are based on a diquark mass of $445$~MeV or a binding
energy of $145$~MeV. Because of the thermal factor, the number of
diquarks decreases as the diquark mass increases. This would reduce
both $\Lambda_c$ and $\Lambda_b$ production and increase that of
$D^0$ and ${\bar B}^0$, reducing thus the $\Lambda_Q/H^0$ ratios.
For example, if the diquark mass is $550$~MeV, the enhancement of
the $\Lambda_c/D^0$ ratio due to diquarks would reduce from 1.6 to
$1.3$. The peak position in the ratio changes, however, very little.

In studying resonance production in the coalescence model, we have
assumed that the effect due to energy mismatch between quarks and
produced hadron is small. Since the coalescence model can be viewed
as formation of bound states from interacting particles in the
system with energy mismatch balanced by other particles in the
system, this would be reasonable if the energy mismatch is small.
Otherwise, the correction factor can, in principle, be determined by
evaluating the transition probability in the presence of other
particles. Also, the coalescence model can be further improved by
imposing energy conservation using Breit-Wigner type spectral
functions for the produced hadrons~\cite{RR07}. This would naturally
lead to suppressed production rate due to the energy mismatch
between quarks and produced hadron. Improving the coalescence model,
therefore, deserves further studies.

One may also consider contributions to heavy hadron production from
other diquark states. It is well-known that the color-spin
interaction for color-sextet diquark of spin-1 is, although not
strong, attractive. (See, e.g., Ref~\cite{HKLW98}.) This diquark, of
course, cannot form a color-singlet baryon with one heavy quark, and
color-sextet diquarks are thus disfavored in phenomenological
models~\cite{Jaffe04}. However, color-sextet diquarks in QGP can
produce color-singlet baryons by first forming color-octet or
color-decuplet baryons and then neutralizing their colors by
interacting with quarks and gluons in QGP. Therefore, if we assume
the existence of (weakly-bound) color-sextet vector diquarks in QGP,
it may bring in additional enhancement of $\Lambda_Q$ baryon
production. Since this mechanism is very model-dependent, the
existence and properties of such diquark states in QGP should be
understood before estimating their contributions to heavy hadron
production.

\acknowledgments

We would like to thank Zi-Wei Lin for help with the HIJING/PYTHIA
simulations. This work was supported by the US National Science
Foundation under Grants No.\ PHY-0457265 and PHY-0758155, the Welch
Foundation under Grant No.\ A-1358, and the Korea Research
Foundation under Grant No.\ KRF-2006-C00011.


\begin{thebibliography}{10}

\bibitem{PHENIX06b}
PHENIX Collaboration, A.~Adare {\em et~al.},
\newblock Phys. Rev. Lett. \textbf{98}, 172301 (2007).
%%CITATION = NUCL-EX 0611018;%%

\bibitem{STAR06c}
STAR Collaboration, B.~I. Abelev,
\newblock Phys. Rev. Lett. \textbf{98}, 192301 (2007).
%%CITATION = NUCL-EX 0607012;%%

\bibitem{DK01b}
\mbox{Yu}. L.~Dokshitzer and D.~E. Kharzeev,
\newblock Phys. Lett. \textbf{B519}, 199 (2001).
%%CITATION = HEP-PH 0106202;%%

\bibitem{DG04}
M.~Djordjevic and M.~Gyulassy,
\newblock Nucl. Phys. \textbf{A733}, 265 (2004).
%%CITATION = NUCL-TH 0310076;%%

\bibitem{VR04}
H.~van Hees and R.~Rapp,
\newblock Phys. Rev. C \textbf{71}, 034907 (2005).
%%CITATION = NUCL-TH 0412015;%%

\bibitem{DGVW05}
M.~Djordjevic, M.~Gyulassy, R.~Vogt, and S.~Wicks,
\newblock Phys. Lett. \textbf{B632}, 81 (2006).
%%CITATION = NUCL-TH 0507019;%%

\bibitem{VGR06}
H.~van Hees, V.~Greco, and R.~Rapp,
\newblock Phys. Rev. C \textbf{73}, 034913 (2006).
%%CITATION = NUCL-TH 0508055;%%

\bibitem{KL07}
C.~M. Ko and W.~Liu,
\newblock Nucl. Phys. \textbf{A783}, 233c (2007).
%%CITATION = NUPHA,A783,233;%%

\bibitem{ZCK05}
B.~Zhang, L.-W. Chen, and C.-M. Ko,
\newblock Phys. Rev. C \textbf{72}, 024906 (2005).
%%CITATION = NUCL-TH 0502056;%%

\bibitem{GAP08}
P.~B. Gossiaux, J.~Aichelin, and A.~Peshier,
arXiv:0802.2525.
%%CITATION = ARXIV:0802.2525;%%

\bibitem{SD05-Soren07}
P.~Sorensen and X.~Dong,
\newblock Phys. Rev. C \textbf{74}, 024902 (2006);
%%CITATION = NUCL-TH 0512042;%%
P.~Sorensen,
\newblock Eur. Phys. J. C \textbf{49}, 379 (2007).
%%CITATION = EPHJA,C49,379;%%

\bibitem{MGC07}
G.~Mart\'{\i}nez-Garc\'{\i}a, S.~Gadrat, and P.~Crochet,
\newblock Phys. Lett. \textbf{B663}, 55 (2008).
%%CITATION = ARXIV:0710.2152;%%

\bibitem{GKL03}
V.~Greco, C.~M. Ko, and P.~L{\'e}vai,
\newblock Phys. Rev. Lett. \textbf{90}, 202302 (2003).
%%CITATION = NUCL-TH 0301093;%%

\bibitem{GKL03b}
V.~Greco, C.~M. Ko, and P.~L{\'e}vai,
\newblock Phys. Rev. C \textbf{68}, 034904 (2003).
%%CITATION = NUCL-TH 0305024;%%

\bibitem{FMNB03}
R.~J. Fries, B.~M\"{u}ller, C.~Nonaka, and S.~A. Bass,
\newblock Phys. Rev. Lett. \textbf{90}, 202303 (2003).
%%CITATION = NUCL-TH 0301087;%%

\bibitem{Khar96}
D.~Kharzeev,
\newblock Phys. Lett. \textbf{B378}, 238 (1996).
%%CITATION = NUCL-TH 9602027;%%

\bibitem{VGW98}
S.~E. Vance, M.~Gyulassy, and X.-N. Wang,
\newblock Phys. Lett. \textbf{B443}, 45 (1998).
%%CITATION = NUCL-TH 9806008;%%

\bibitem{VG99}
S.~E. Vance and M.~Gyulassy,
\newblock Phys. Rev. Lett. \textbf{83}, 1735 (1999).
%%CITATION = NUCL-TH 9901009;%%

\bibitem{TGBG05}
V.~Topor~Pop, M.~Gyulassy, J.~Barrette, and C.~Gale,
\newblock Phys. Rev. C \textbf{72}, 054901 (2005).
%%CITATION = HEP-PH 0505210;%%

\bibitem{LOYYK07}
S.~H. Lee, K.~Ohnishi, S.~Yasui, I.-K. Yoo, and C.~M. Ko,
\newblock Phys. Rev. Lett. \textbf{100}, 222301 (2008).
%%CITATION = ARXIV:9709.3637;%%

\bibitem{SZ03b-SZ04}
E.~V. Shuryak and I.~Zahed,
\newblock Phys. Rev. C \textbf{70}, 021901(R) (2004);
%%CITATION = HEP-PH 0307267;%%
\newblock Phys. Rev. D \textbf{70}, 054507 (2004).
%%CITATION = HEP-PH 0403127;%%

\bibitem{DS88}
J.~F. Donoghue and K.~S. Sateesh,
\newblock Phys. Rev. D \textbf{38}, 360 (1988).
%%CITATION = PHRVA,D38,360;%%

\bibitem{Sateesh92}
K.~S. Sateesh,
\newblock Phys. Rev. D \textbf{45}, 866 (1992).
%%CITATION = PHRVA,D45,866;%%

\bibitem{IK66}
M.~Ida and R.~Kobayashi,
\newblock Prog. Theor. Phys. \textbf{36}, 846 (1966).
%%CITATION = PTPKA,36,846;%%

\bibitem{LT67-LTK68}
D.~B. Lichtenberg and L.~J. Tassie,
\newblock Phys. Rev. \textbf{155}, 1601 (1967);
%%CITATION = PHRVA,155,1601;%%
D.~B. Lichtenberg, L.~J. Tassie, and P.~J. Keleman,
\newblock Phys. Rev. \textbf{167}, 1535 (1968).
%%CITATION = PHRVA,167,1535;%%

\bibitem{SV05}
M.~Shifman and A.~Vainshtein,
\newblock Phys. Rev. D \textbf{71}, 074010 (2005).
%%CITATION = HEP-PH 0501200;%%

\bibitem{APEFL93}
M.~Anselmino {\em et~al.},
\newblock Rev. Mod. Phys. \textbf{65}, 1199 (1993).
%%CITATION = RMPHA,65,1199;%%

\bibitem{BDTR02}
A.~Bender, W.~Detmold, A.~W. Thomas, and C.~D. Roberts,
\newblock Phys. Rev. C \textbf{65}, 065203 (2002).
%%CITATION = NUCL-TH 0202082;%%

\bibitem{SW06}
A.~Selem and F.~Wilczek, arXiv:hep-ph/0602128.
%%CITATION = HEP-PH 0602128;%%

\bibitem{EFJL83}
S.~Ekelin, S.~Fredriksson, M.~J{\"a}ndel, and T.~I. Larsson,
\newblock Phys. Rev. D \textbf{28}, 257 (1983).
%%CITATION = PHRVA,D28,257;%%

\bibitem{Jaffe04}
R.~L. Jaffe,
\newblock Phys. Rep. \textbf{409}, 1 (2005).
%%CITATION = HEP-PH 0409065;%%

\bibitem{ADL06}
C.~Alexandrou, {\mbox{Ph}}.~de~Forcrand, and B.~Lucini,
\newblock Phys. Rev. Lett. \textbf{97}, 222002 (2006).
%%CITATION = HEP-LAT 0609004;%%

\bibitem{HKLW98}
M.~Hess, F.~Karsch, E.~Laermann, and I.~Wetzorke,
\newblock Phys. Rev. D \textbf{58}, 111502 (1998).
%%CITATION = HEP-LAT 9804023;%%

\bibitem{LNW83}
D.~B. Lichtenberg, W.~Namgung, and J.~G. Wills,
\newblock Z. Phys. C \textbf{19}, 19 (1983).
%%CITATION = ZEPYA,C19,19;%%

\bibitem{EFKR96}
D.~Ebert, T.~Feldmann, C.~Kettner, and H.~Reinhardt,
\newblock Z. Phys. C \textbf{71}, 329 (1996).
%%CITATION = HEP-PH 9506298;%%

\bibitem{FSR88}
S.~Fleck, B.~Silvestre-Brac, and J.~M. Richard,
\newblock Phys. Rev. D \textbf{38}, 1519 (1988).
%%CITATION = PHRVA,D38,1519;%%

\bibitem{HNV08}
E.~Hern{\'a}ndez, J.~Nieves, and J.~M. Verde-Velasco,
\newblock Phys. Lett. \textbf{B666}, 150 (2008).
%%CITATION = ARXIV: 0803.3672;%%

\bibitem{SMS06}
T.~Sj{\"o}strand, S.~Mrenna, and P.~Skands,
JHEP \textbf{0605}, 026 (2006).
%%CITATION = HEP-PH 0603175;%%

\bibitem{PDG08}
Particle Data Group, C.~Amsler {\em et~al.},
\newblock Phys. Lett. \textbf{B667}, 1 (2008),
\newblock http://pdg.lbl.gov.
%%CITATION = PHLTA,B667,1;%%

\bibitem{SELEX00}
SELEX Collaboration, A.~Kushnirenko {\em et~al.},
\newblock Phys. Rev. Lett. \textbf{86}, 5243 (2001).
%%CITATION = HEP-EX 0010014;%%

\bibitem{ZKL07}
B.-W. Zhang, C.~M. Ko, and W.~Liu,
\newblock Phys. Rev. C \textbf{77}, 024901 (2007).
%%CITATION = ARXIV:0709.1684;%%

\bibitem{KR06b}
I.~Kuznetsova and J.~Rafelski,
\newblock Eur. Phys. J. C \textbf{51}, 113 (2007).
%%CITATION = HEP-PH 0607203;%%

\bibitem{ABRS07}
A.~Andronic, P.~Braun-Munzinger, K.~Redlich, and J.~Stachel,
\newblock Phys. Lett. \textbf{B659}, 149 (2008).
%%CITATION = ARXIV:0708.1488;%%

\bibitem{OPM94c-OP96a}
Y.~Oh, B.-Y. Park, and D.-P. Min,
\newblock Phys. Rev. D \textbf{49}, 4649 (1994);
%%CITATION = HEP-PH/9402205;
\newblock Phys. Rev. D \textbf{50}, 3350 (1994);
%%CITATION = HEP-PH 9407214;%%
Y.~Oh and B.-Y. Park,
\newblock Phys. Rev. D \textbf{53}, 1605 (1996).
%%CITATION = HEP-PH 9510268;%%

\bibitem{CDF07}
CDF Collaboration, T.~Aaltonen {\em et~al.},
\newblock Phys. Rev. Lett. \textbf{99}, 202001 (2007).
%%CITATION = ARXIV 0706.3868;%%

\bibitem{KKLR08}
M.~Karliner, B. Keren-Zur, H.~J. Lipkin, and J.~L. Rosner,
arXiv:0804.1575.
%%CITATION = 0804.1575;%%

\bibitem{CK06}
L.-W. Chen and C.~M. Ko,
\newblock Phys. Rev. C \textbf{73}, 044903 (2006).
%%CITATION = NUCL-TH 0602025;%%

\bibitem{GKR04}
V.~Greco, C.~M. Ko, and R.~Rapp,
\newblock Phys. Lett. \textbf{B595}, 202 (2004).
%%CITATION = NUCL-TH 0312100;%%

\bibitem{Hwang02}
C.-W. Hwang,
\newblock Eur. Phys. J. C \textbf{23}, 585 (2002).
%%CITATION = HEP-PH 0112237;%%

\bibitem{brac96} 
B.~Silvestre-Brac, 
Few-Body Syst. \textbf{20}, 1 (1996).
%%CITATION = FBSYE,20,1;%%

\bibitem{CKLN07}
L.~W. Chen, C.~M. Ko, W.~Liu, and M.~Nielsen,
\newblock Phys. Rev. C \textbf{76}, 014906 (2007).
%%CITATION = ARXIV:0705.1697;%%

\bibitem{BD96a}
A.~J. Baltz and C.~Dover,
\newblock Phys. Rev. C \textbf{53}, 362 (1996).
%%CITATION = PHRVA,C53,362;%%

\bibitem{OK07}
Y.~Oh and C.~M. Ko,
\newblock Phys. Rev. C \textbf{76}, 054910 (2007).
%%CITATION = ARXIV:0707.3332;%%

\bibitem{CGKL03}
L.~W. Chen, V.~Greco, C.~M. Ko, S.~H. Lee, and W.~Liu,
Phys. Lett. \textbf{B601}, 34 (2004).
%%CITATION = nucl-th/0308006;%%

\bibitem{CRP87}
R.~T. Cahill, C.~D. Roberts, and J.~Praschifka,
\newblock Phys. Rev. D \textbf{36}, 2804 (1987).
%%CITATION = PHRVA,D36,2804;%%

\bibitem{CNV05}
M.~Cacciari, P.~Nason, and R.~Vogt,
\newblock Phys. Rev. Lett.  \textbf{95}, 122001 (2005).
%%CITATION = HEP-PH 0502203;%%

\bibitem{STAR04d}
STAR Collaboration, J.~Adams {\em et~al.},
\newblock Phys. Rev. Lett. \textbf{94}, 062301 (2005).
%%CITATION = NUCL-EX 0407006;%%

\bibitem{LTTW97b}
D.~H. Lu, K.~Tsushima, A.~W. Thomas, A.~G. Williams, and K.~Saito,
Nucl. Phys. \textbf{A634}, 443 (1998);
%%CITATION = NUCL-TH 9712027;%%
X.~Jin and B.~K. Jennings,
Phys. Rev. C \textbf{54}, 1427 (1996).
%%CITATION = NUCL-TH 0604018;%%

\bibitem{KM06b}
Y. Kanada-En'yo and B. M{\"u}ller,
\newblock Phys. Rev. C \textbf{74}, 061901(R) (2006).
%%CITATION = NUCL-TH 0608015;%%

\bibitem{PHENIX04b}
PHENIX Collaboration, S.~S. Adler {\em et~al.},
\newblock Phys. Rev. C \textbf{71}, 051902(R) (2005).
%%CITATION = NUCL-EX 0408007;%%

\bibitem{STAR06d}
STAR Collaboration, J.~Adams {\em et~al.}, arXiv:nucl-ex/0601042.
%%CITATION = NUCL-EX 0601042;%%

\bibitem{Soren06}
P.~Sorensen,
\newblock J. Phys. G \textbf{32}, S135 (2006).
%%CITATION = NUCL-EX 0701048;%%

\bibitem{ZKLL00a}
B.~Zhang, C.~M. Ko, B.-A. Li, and Z.~Lin,
\newblock Phys. Rev. C \textbf{61}, 067901 (2000).
%%CITATION = NUCL-TH 9907017;%%

\bibitem{LKLZP04}
Z.-W. Lin, C.~M. Ko, B.~A. Li, S.~Pal, and B.~Zhang,
\newblock Phys. Rev. C \textbf{72}, 064901 (2005).
%%CITATION = NUCL-TH 0411110;%%

\bibitem{LK02}
W.~Liu and C.~M. Ko,
\newblock Phys. Lett. \textbf{B533}, 259 (2002);
%%CITATION = NUCL-TH 0201074;%%
W.~Liu, C.~M. Ko, and S.~H. Lee,
\newblock Nucl. Phys. {\bf A728}, 457 (2003).
%%CITATION = NUCL-TH 0308013;%%

\bibitem{lin}
Z.~Lin, C.~M. Ko, and B. Zhang,
\newblock Phys. Rev. C {\bf 61}, 024904 (2000);
%%CITATION = NUCL-TH/9905003;%%
Z.~Lin, T.~G. Di, and C.~M. Ko,
\newblock Nucl. Phys. {\bf A689}, 965 (2001).
%%CITATION = NUCL-TH/0006086;%%

\bibitem{RR07}
L. Ravagli and R. Rapp, Phys. Lett. {\bf B655}, 126 (2007).
%%CITATION = ARXIV:0705.0021;%%

\end{thebibliography}
\end{document}